\documentclass[letterpaper]{article}
\usepackage{aaai17}
\usepackage{times}
\usepackage{helvet}
\usepackage{courier}
\usepackage{enumitem}
\usepackage{graphicx}
\usepackage{tabu}
\usepackage{booktabs}
\usepackage{url}
\usepackage{flushend}
\usepackage{subcaption}
\usepackage{color}
\usepackage{multirow}
\usepackage{rotating}
\graphicspath{ {images/} }
\newcommand{\ifrac}[2]{%
	\mathchoice
	{\frac{#1}{#2}}
	{#1/#2}
	{#1/#2}
	{#1/#2}
}

\title{It's Always April Fools' Day! On the Difficulty of Social Network Misinformation Classification via Propagation Features}
\author{Mauro Conti, Daniele Lain,\\ {\Large \textbf{Riccardo Lazzeretti}, \textbf{Giulio Lovisotto}}\\
University of Padua\\
\And
Walter Quattrociocchi\\
IMT School for Advanced Studies, Lucca\\}
\begin{document}
\maketitle
\begin{abstract}


Given the huge impact that Online Social Networks (OSN) had in the way people get informed and form their opinion, they became an attractive playground for malicious entities that want to spread misinformation, and leverage their effect.
In fact, misinformation easily spreads on OSN and is a huge threat for modern society, possibly influencing also the outcome of elections, or even putting people's life at risk (e.g.,  spreading ``anti-vaccines'' misinformation). 
Therefore, it is of paramount importance for our society to have some sort of ``validation'' on information spreading through OSN. The need for a wide-scale validation would greatly benefit from automatic tools.

In this paper, we show that it is difficult to carry out an automatic classification of misinformation considering only structural properties of content propagation cascades. We focus on structural properties, because they would be inherently difficult to be manipulated, with the the aim of circumventing classification systems.
To support our claim, we carry out an extensive evaluation on Facebook posts belonging to conspiracy theories 
(as representative of misinformation), and scientific news (representative of fact-checked content).
Our findings show that conspiracy content actually reverberates in a way which is hard to distinguish from the one scientific content does: for the classification mechanisms we investigated, classification $F_1$-score never exceeds 0.65 during content propagation stages, and is still less than 0.7 even after propagation is complete.
\end{abstract}
\section{Introduction}
\label{sec:introduction}


An increasing number of people get informed on Online Social Networks (OSN)~\cite{newman2015reuters}.
However, as OSN allow every user to post content, which propagates among users through viral processes, these platforms became attractive targets for misinformation creators.
Moreover, the hyperconnected world and increasing complexity of reality create a scenario in which viral processes on OSN are driven by confirmation bias, eliciting the proliferation of unsubstantiated rumors and hoaxes all the way up to conspiracy theories \cite{bessi2015viral,bessi2014science}.
News stories undergo the same popularity dynamics as other forms of online contents (such as selfies and cat pictures)~\cite{newman2015reuters}.
It is not a surprise then that the Oxford Dictionary in 2016 elected \textit{Post-truth} as word of the year.
The definition reads:
\begin{quote}
\textit{``Relating to or denoting circumstances in which objective facts are less influential in shaping public opinion than appeals to emotion and personal belief''.}
\end{quote}

Several studies pointed out the effects of social influence online~\cite{centola2010,fowler2010cooperative,quattrociocchi2014opinion,salganik2006experimental}.
Results reported in \cite{kramer2014} indicate that emotions expressed by others on Facebook influence our own emotions, providing experimental evidence of massive-scale contagion via social networks.
As a result of disintermediated access to information and of algorithms used in content promotion, communication has become increasingly personalized, both in the way messages are framed and how they are shared across social networks. 
Selective exposure has been shown to favor the emergence of echo-chambers --- polarized groups of like-minded people where users reinforce their world view with information adhering to their system of beliefs~\cite{del2016spreading}.
Confirmation bias, indeed, plays a pivotal role in informational cascades~\cite{quattrociocchi2016echo,bessi2015viral,zollo2015debunking,sunstein2002law}.
Recent works~\cite{bessi2014science,zollo2015debunking} showed that attempts to debunk false information are largely ineffective.
In particular, discussion degenerates when the two polarized communities interact with one another~\cite{zollo2015emotional}.
OSN users therefore tend to select information that are consistent with their beliefs (even if containing false claims) and propagate it to like-minded friends, and to ignore information dissenting with their beliefs.
This confirms that misinformation is a huge threat for modern society.
Not only it can put people's life at risk, as in the case of ``anti-vaccines'' misinformation, it is also starting to be used against political opponents, such as in the US, during the 2016 electoral campaign~\cite{buzzfeedhoaxes} and during Election Day~\cite{electiondayshoaxes}.

Rising attention to the spreading of fake news and unsubstantiated rumors led researchers to investigate many of their aspects, from the characterization of conversation threads~\cite{backstrom2013characterizing}, to the detection of bursty topics on microblogging platforms~\cite{diao2012finding}, to the disclosure of the mechanisms behind information diffusion for different kinds of contents~\cite{romero2011differences}.
Spreading of misinformation also motivated major corporations like Google and Facebook to provide solutions to the problem~\cite{firstdraft}.
However, given the amount of content posted everyday on OSN (e.g., Facebook reports 1.18 billion daily active users on September 2016~\cite{fbnewsroom}), effective fact-checking would greatly benefit of automatic classification tools, that would possibly not require human intervention.
Moreover, classification of fake news and misinformation should ideally use properties that misinformation creators can not manipulate.
Considering in the classification, for example, the trustworthiness of news domains, or the topic of content, could lead to an arms race with creators of false news.
Instead, the topology of propagation cascades, and the patterns of users' interaction with content, are outside of the domain of our ``adversaries'', and are much more difficult to be manipulated.

In this work, we investigate detection of viral processes by comparing diffusion of posts from scientific and conspiracy pages on the Italian Facebook network. The former diffuse scientific knowledge, where details about the sources (such as authors and funding programs) are easy to access. We therefore select their posts as representative of fact-checked content. The latter aim at diffusing what is neglected by ``manipulated'' mainstream media. Specifically, conspiracy theories tend to reduce the complexity of reality by explaining significant social or political aspects as plots conceived by powerful individuals or organizations. Since these kinds of arguments can sometimes involve the rejection of science, alternative explanations are invoked to replace the scientific evidence. For instance, people who reject the link between HIV and AIDS generally believe that AIDS was created by the US Government to control the African American population. We therefore select their posts as representative of misinformation.




\paragraph{Contribution.}
The contributions of this paper are the following:
\begin{itemize}
    \item We show that automatic fact-checking with classification techniques employing only structural features of content propagation cascades (features that are robust to attacker's manipulation) does not suggest to bring usable results. Given the grain of our data, we design classifiers that leverages topological properties of content propagation cascades, and properties of the evolution over time of users' interactions with content. A set of classifiers operates with evolution properties to classify content during its early propagation stage, another set of classifiers operates with more features to classify content that already propagated. Indeed, being able to issue warnings about possible fake news as early as possible, and retroactively flag such news can be useful, in the fight against OSN misinformation. 
    \item We evaluate our classifiers on a well-known dataset of Facebook posts from Italian pages. We use posts belonging to conspiracy theories as representative of misinformation, and posts belonging to scientific news as representative of fact-checked content. Our findings highlight the complexity of creating automatic solutions to misinformation classification. Indeed, structural features of content propagation do not allow us to obtain notable improvements from a random guess baseline: $F_1$-score for the classification mechanisms we investigated is always lower than 0.65 during content propagation stages, and is still less than 0.7 even after propagation is complete.
\end{itemize}

\paragraph{Outline.}
Section~\ref{sec:related_work} overviews related work in news and misinformation propagation in OSN. Section~\ref{sec:methods} formally models and defines content propagation in Facebook. Section~\ref{sec:experiments} presents our experiment design and the features we extract from content propagation cascades. Then, Section~\ref{sec:results} evaluates our classification and, finally, Section~\ref{sec:conclusions} critically discusses our results, summarizes the paper and delineates future work.

\section{Related Work}
\label{sec:related_work}
Several studies moved towards the spreading of rumors and behaviors on online social networks, challenging both their structural properties and their effects on social dynamics \cite{moreno2004dynamics,doerr2012rumors,seo2012identifying,borge2013emergence,cozzo2013contact,borge2012locating}. 
 In \cite{ugander2012}, authors find that the probability of contagion is tightly controlled by the number of connected components in an individual's contacts neighborhood, rather than by the actual size of the neighborhood. In \cite{centola2007} researchers show that, although long ties are relevant for spreading information about an innovation or social movement, they are not sufficient with respect to the social reinforcement necessary to act on that information. 
A key factor in identifying true contagion in social network is to distinguish between peer-to-peer influence and homophily: in the first case, a node influences or causes outcomes to its neighbors, whereas in the second one, dyadic similarities between nodes create correlated outcome patterns among neighbors that could mimic viral contagions even without direct causal influence \cite{mcpherson2001}. 
The study presented in \cite{aiello2012friendship} reveals that there is a substantial level of topical similarity among users which are close to each other in the social network, suggesting that users with similar interests are more likely to be friends. 
In \cite{aral2009} authors develop an estimation framework to distinguish influence and homophily effects in dynamic networks and find that homophily explains more than 50\% of the perceived behavioral contagion. In \cite{bakshy2012} the analysis faces the role of social networks and exposure to friends' activities in information resharing on Facebook. Once having isolated contagion from other confounding effects such as homophily, authors claim that there is a considerably higher chance to share contents when users are exposed to friends' resharing. 
All these contributions strive to understand the inner mechanism of rumor spreading and to eventually predict massive diffusion processes, i.e. cascades.
Cascades recurrence and prediction has been shaped in \cite{cheng2014can} and \cite{cheng2016cascades}.

\section{Methods}
\label{sec:methods}
In this section, we first report and describe the employed dataset (Section~\ref{sec:dataset}). We then give some necessary background knowledge, and present our reference model and definition of content propagation mechanisms of Facebook (Section~\ref{sec:background}).

\subsection{Dataset}\label{sec:dataset}
We employ a well-known dataset of posts shared by Italian Facebook users~\cite{bessi2014science}.
This dataset contains posts published by 73 public Facebook pages: 34 pages that publish scientific content (e.g., press releases of peer-reviewed articles), and 39 pages that publish conspiracy theories-related content (e.g. new world order, chemtrails).
Additionally, the dataset contains information about the interaction of users with these posts, and users' ego-networks (i.e., the list of users that are their friends, when such list is public).
Additionally, for a set of posts, the dataset provides information about the \textit{propagation cascade} of such content, generated by users' reshares, and subsequent reshares from their friends. This propagation from one user to other users can happen multiple times, forming a cascade of resharing.
Using information from the dataset, we extracted 112141 non-empty propagation cascades, 89491 for conspiracy and 22650 for science, respectively. We underline that the dataset is obtained by using the Facebook Graph API, and contains only public information. Hence, timestamps of \textit{reshares} and \textit{comments} are available, but timestamps of \textit{like} interactions are not.

\subsection{Background and Definitions}\label{sec:background}
We now present our reference formal representation of Facebook's \textit{friendship graph}, and the \textit{potential propagation graph} generated by content posted on the social network.

\paragraph{Facebook Friendship Graph.}
We model Facebook relationships as a graph $\mathcal{G}\langle V,E\rangle$, that we call \textit{Facebook friendship graph}, where $V$ is the set of nodes that represent \textit{entities}, namely user accounts and page accounts. We assume two main differences among these two types of entity: (i) pages can post new content on the OSN, while users can only interact with such content by liking, commenting, and resharing it; and (ii) users can establish \textit{friendship} relationships with other users, while pages cannot. Indeed, two users $v_1, v_2 \in V$ are connected by an edge $e(v_1, v_2) \in E$ if they are \textit{friends} on Facebook. Pages are not connected by any edge, as they do not have proper friendship relationships. This model is a simplification of how Facebook actually works, because users can post new content, and pages and users are linked by \textit{like} relationships.
Similarly, users can \textit{follow} other users, without having any friend relationship with them. 
However, for this work, we do not focus on new content generated by users. Moreover, the dataset we use lacks information about the \textit{like} and \textit{follow} relationships, that we therefore can not consider.

\paragraph{Potential Propagation Graph.}
Before formally modeling the spreading of content on Facebook, we describe some fundamental concepts of the OSN.
We recall that, in our model, only pages can post new content on the social network.
Henceforth, we refer to the page that originally posted some content as the \textit{seed page}. Instead, users find new content by looking at their \textit{timeline}, where they see recent content posted by the pages they like, and content that their friends recently interacted with.
They can then interact with such content through the means of \textit{resharing}, \textit{liking}, and \textit{commenting} it.
However, all of these interactions can happen directly on the original content, or on some types of interactions of users' friends.
For example, a user observing a comment or reshare of a post by one of his friends can decide to interact directly with the original post, or with his friend's interaction itself.
In the first case, user's interaction looks exactly the same and it is impossible to understand whether the content was found thanks to his friend's interaction, or directly on the seed page.
In fact, differently from related work~\cite{friggeri2014rumor}, the dataset we employ does not provide this type of information.
We therefore take a conservative approach, saying that the content \textit{potentially propagated} to the user both from the seed page and from any of his friends that interacted with the content, without any distinction.
On the other hand, in the second case, it becomes clear that the user found the content thanks to his friend. We therefore say that the content \textit{propagated} to the user from his friend.

We now formalize the above observations.
Let $P$ be the set of contents posted on Facebook by seed pages.
For each post $p \in P$, created at some time $t$ by a seed page, at a generic subsequent point in time ${t+\delta}$ we define a \textit{potential propagation graph} ~$ \mathcal{G}^p_{t+\delta}\langle V^p_{t+\delta}, E^p_{t+\delta}\rangle $, where $V^p_{t+\delta}$ is the set composed by the seed page, and by the users that interacted with $p$ during the time interval $[t, {t+\delta}]$.
\textit{Final potential propagation graph} $\mathcal{G}^p\langle V^p, E^p\rangle$ is the graph formed considering all interactions with $p$ on the timespan of the analysis (as new interactions with old content are always possible on Facebook).
Two nodes $v_1, v_2 \in V^p_{t+\delta}$ are connected by an undirected \textit{potential propagation edge} $e^p(v_1, v_2) \in E^p_{t+\delta}$ if either (i) $v_1$ or $v_2$ already interacted with $p$ and $\exists e\:|\:e(v_1, v_2) \in E$ (that is, $v_1$ and $v_2$ are friends on Facebook), or (ii) $v_1$ or $v_2$ is the seed page.
Therefore, an edge $e(v_1, v_2) \in E^p_{t+\delta}$ indicates that the content $p$ \textit{potentially propagated} either from $v_1$ to $v_2$, or from $v_2$ to $v_1$.
We associate to each edge $e \in E^p_{t+\delta}$ two different properties:
\begin{enumerate}[leftmargin=.5in]
\item \textbf{Time}: the time when the interaction between $v_1$ and $v_2$ took place;
\item \textbf{Type}: either ``like'', ``comment'', ``reshare'', or ''friendship'', depending on the type of interaction between $v_1$ and $v_2$.
\end{enumerate}
Hereafter in the paper, we will refer to these properties for an edge $e \in E^p_{t+\delta}$ with $e.property\_name\:$ (for example, $e.time$).

\begin{figure}[h]
	\centering
	\begin{subfigure}[b]{0.48\linewidth}
	\includegraphics[width=\textwidth]{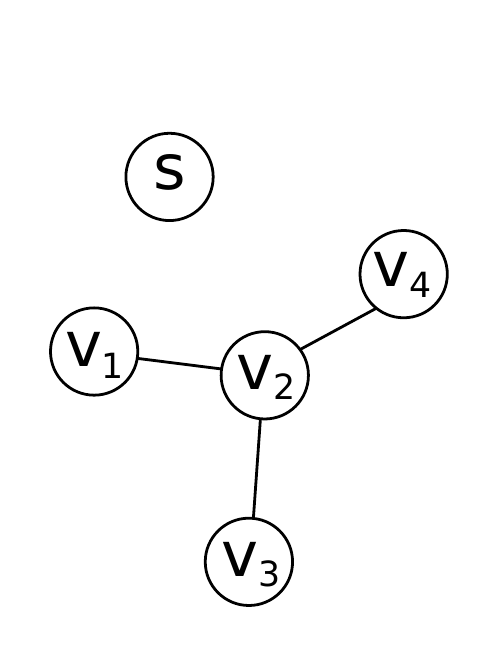}
        \caption{Facebook Friendship Graph}
        \label{fig:graph_e1}
	\end{subfigure}
	~
	\begin{subfigure}[b]{0.48\linewidth}
	\includegraphics[width=\textwidth]{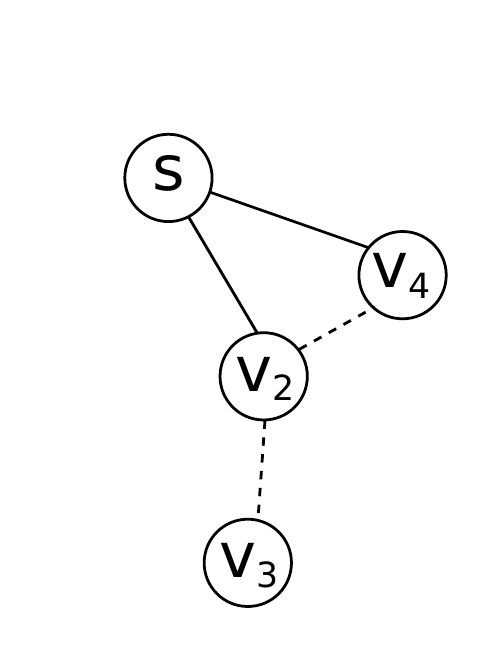}
        \caption{Potential Propagation Graph}
        \label{fig:graph_e2}
	\end{subfigure}
	\caption{Sample Facebook friendship graph, and potential propagation graph of some content. A dashed line represents potential propagation edges that also corresponds to a friendship edge of the friendship graph. Node $v1$ does not interact with the content, and is therefore not present in the potential propagation graph.}
	\label{fig:edges}
\end{figure}

Figure~\ref{fig:edges} depicts an example of this propagation model. We represent a simple Facebook friendship graph in Figure~\ref{fig:graph_e1}, where edges represent friendship relations. Nodes $v_1, ..., v_4$ are OSN users, while node $s$ represents a public page posting content on the platform. We suppose $v_2$ and $v_4$ interacted with a given content posted by $s$, and $v_3$ reshared the post from $v_2$. Node $v_1$ did not interact with the content, hence is not present on the potential propagation graph of the post, that we represent in Figure~\ref{fig:graph_e2}. There, with edges $(s, v_2), (s, v_4), (v_2, v_4)$ we represent possible propagation paths of the content: either $v_2$ or $v_4$ could have seen the content from the seed page, or from previous interaction of their friend. Node $v_3$ only has an edge $(v_2, v_3)$, as we know that the interaction happened thanks to $v_2$, and therefore the content propagated from this node. Additionally, we highlight that some possible propagation edges, represented with a dashed line, correspond to friendship edges of the friendship graph.

\section{Experiments}
\label{sec:experiments}
In this section, we describe the details of the experiment and of the analysis performed on the dataset. We first describe the experimental design (Section~\ref{sec:experimental_design}). We then thoroughly report and motivate the different features that we extracted from propagation cascades (Section~\ref{sec:features}).

\subsection{Experimental Design}\label{sec:experimental_design}
The aim of our experiments is to show that it is difficult to discriminate between conspiracy theories, and fact-checked scientific news, by using only the propagation graph of the post, as it would be difficult to manipulate by misinformation creators. In particular, we evaluate this in two specific moments of content propagation:
\begin{enumerate}[leftmargin=.5in]
    \item \textbf{Early Stage}, meaning that classification of the type of post happens as early as possible during its propagation phase;
    \item \textbf{Final Stage}, meaning that classification happens when the post already stopped propagating (within the considered timespan of the employed dataset), and its cascade is complete. 
\end{enumerate}
These two scenarios are of particular interest, both from a research and a practical points of view. Indeed, being able to issue automated warnings as early as possible about potential misinformation could help in reducing their spread~\cite{friggeri2014rumor}. 
On the other hand, still being able to perform such a classification when the post already propagated might serve as a warning in order to prevent its further diffusion. Moreover, retroactively flagging old posts as potentially fake could help training users to discriminate between fact-checked and dubious information, a major direction in the fight against misinformation~\cite{firstdraft}.

To investigate these two scenarios, we set up two different experiments, modeled as binary classification tasks, where the two classes are \textit{conspiracy} and \textit{science}. We describe them in the following.

\paragraph{Early Stage Classification.} In this scenario, we have access only to users' interactions with a post up to a certain time $t+\delta$, after it has been created at time $t$, (e.g., up to after two hours after creation). With this scenario, we simulate how OSNs such as Facebook could try to continuously detect misinformation content during its propagation. Therefore, for each post $p$, given its creation time $t$, we extract features from the partial propagation graph $\mathcal{G}^p_{t+\delta}$, and from the evolution of properties of the propagation graph before the current time $t+\delta$ (discussed further in Section~\ref{sec:features}). We use 30 minutes steps, as this proved to be a good trade-off between the granularity of the analysis and the number of intervals to consider. We stop at two days (2880 minutes, or 96 time steps) after the publication time of a post, because we observed that most of the interactions happen in this period ($> 95 \%$). Finally, we compare the performance of different well-known classifiers, namely Random Forests ~\cite{ho1995}, Linear Discriminant Analysis~\cite{izenman2013linear}, and Multi-Layer Perceptron~\cite{haykin2004comprehensive}, for each $\delta \in [30, 60, ... , 2880]$, in a cross-validation scheme.

\paragraph{Final Stage Classification.} In this scenario, we suppose that a post already stopped its propagation, and no new interactions with it have been observed for a long time. We describe the final propagation graph $\mathcal{G}^p$ of post $p$ with a set of high-level and topological features, that we describe in more detail in Section~\ref{sec:features}. We then compare the performance of different classifiers, in a cross-validation scheme.

\subsection{Feature Extraction}\label{sec:features}
To extract information from the different propagation graphs $\mathcal{G}^p$ and $\mathcal{G}_t^p$, we identify three possible categories of features: (i) high-level properties of the content propagation, (ii) topological properties of the propagation graphs, (iii) evolution properties of the content propagation. We use feature sets (i), (ii), and (iii) considering the evolution after two days, in the Final Stage Classification scenario. We only use feature set (iii), in the Early Stage Classification scenario, because the other features characterize only $\mathcal{G}^p$, but not $\mathcal{G}_t^p$. We summarize these different features in Table~\ref{tab:features}, along with their formal definitions, and describe them in the following.

{
\tabulinesep = 1.5mm
\begin{table*}[!t]
\begin{tabu} to \textwidth {X[1.2]X[4]}
\tabucline-
\rowfont[c] \bfseries Feature Name & \hspace{-4.2cm} \textbf{Description}\\
\hline\hline 
\multicolumn{2}{c}{ \textbf{High Level Properties}}\\
Size & Number of edges of the graph $|E^p|$\\ 
Friendships Ratio & Proportion of edges whose type is ``friendship'': $\ifrac{|\{e \in E^p \: | \: e.type = ``friendship"\}|}{|E^p|}$ \\
Interactions Ratio & Number of vertices over the number of edges: $\ifrac{|V^p|}{|E^p|}$\\
Lifetime & Time passed between post creation time and the last interaction time: $\max_{e \in E^p}\: e.time - t$\\
90\% Interactions Time& Time required for the content to reach its 90\% number of interactions\\
\tabucline-
\multicolumn{2}{c}{ \textbf{Topological Properties}}\\
Average Vertex Degree & Average possible propagation paths of the post: $\ifrac{2 \cdot |E^p|}{|V^p|}$\\
Clustering Coefficient & Ratio of connected triplets of nodes\\
Assortativity Coefficient& Degree correlation between pairs of linked nodes\\
Average Path Length & Average length of the shortest paths\\
Diameter & Length of the longest shortest path between any pair of vertices\\
\tabucline-
\multicolumn{2}{c}{ \textbf{Evolution Properties}}\\
Mean & $\frac{1}{n}\sum_{i}^{n}v_i$\\
Linear Weighted Mean & $\frac{2}{n(n+1)}\sum_{i}^{n}i \cdot v_i$\\
Quadratic Weighted Mean & $\frac{6}{n(n+1)(2n+1)}\sum_{i}^{n}i^2 \cdot v_i$\\
Standard Deviation & $\sqrt{\frac{\sum_{i}^{n}(v_i - \bar{v})^2}{n}}$, where $\bar{v}$ is the mean of $v$\\
Average Absolute Change & $\frac{1}{n}\sum_{i}^{n-1}|v_i - v_{i+1}|$\\
Maximum & $\max\limits_{i} \: v_i$\\
\tabucline-
\tabuphantomline
\end{tabu}
\caption{Description and formal definition of features we use on our Final Stage and Early Stage classification experiments.}\label{tab:features}
\end{table*}
}

\subsubsection{High Level Properties.}
These features represent high-level properties of the complete propagation cascade represented by $\mathcal{G}^p\langle V^p, E^p\rangle$. 

Some features represent very general properties related to the virality of the content. These general properties are the \textit{lifetime} of the cascade, measured as the distance in minutes from the first to the last captured interaction with the content; the \textit{size} of the cascade in terms of number of nodes (users who interacted with the content); the \textit{number of total interactions}; and the time required for the cascade to reach its 90\% total interactions (referred to as \textit{90\% interactions time}).

Other features derived from high level properties attempt to capture different possible types of interaction with the content. \textit{Friendships ratio} is defined as the proportion of edges whose type is ``friendship'' over the total number of edges and represents the number of times, in proportion, that the post potentially propagated among friends, rather than directly from the seed node. Indeed, if no friends of users interact with some content, its potential propagation graph only contains edges with the seed page. \textit{Interactions ratio}, instead, represents the average exposure to interactions from friends of users with the content. Since vertices are interacting users, and edges are potential interactions, higher values of this metric mean lower exposure (little interaction with the content by one's friends). These features are motivated by the observation that it is possible that the users' fruition of different types of content is different, with some types of content being interacted directly from the source, and other types of content relying more on word-of-mouth propagation~\cite{romero2011differences}.

\subsubsection{Topological Properties.}
We also select as features some well-known properties of the topological structure of graphs. These properties are commonly used to learn information about graph structures, and have been applied in solving problems such as link analysis and prediction~\cite{al2006link}, and especially in cascade and virality prediction~\cite{hong2011predicting,cheng2014can}.
The \textit{average vertex degree} feature represents the average number of possible propagation edges for the content at any given hop. Higher values of this metric indicate the presence of interacting users greatly exposed to the content, or able to influence many of their social friends. The \textit{global clustering coefficient}~\cite{Holland1971}, a measure of the density of connections of graphs, is another indication of whether the possible propagation paths are generated from interactions between friends, or directly with the seed page. \textit{Assortativity coefficient}, defined as the degree correlation between pairs of linked nodes~\cite{newman2002assortative}, can measure how friends influence each others in interacting with content on the social network. \textit{Average path length}~\cite{fronczak2004average}, also known as Wiener index, gives us indications of the virality of the content, in terms of distance of propagation from the seed  page. Long cascading news, reshared many times from interacting friends, will exhibit a longer average path length than news whose interactions happened mostly from the seed node. Finally, the \textit{diameter} of a graph, defined as the longer shortest path between any pair of nodes of the graph, indicates the spreading distance of posts.

\subsubsection{Evolution Properties.}
These features represent evolution properties over time of the post $p$ propagation, from its creation time $t$ to a subsequent point $t+\delta$.
To compute these features, we construct the propagation graphs at different time steps $\mathcal{G}_{t+30},...,\mathcal{G}_{t+\delta}$. We then calculate the value of three of our high-level features for each graph at each time step: (i) Friendships Ratio, (ii) Size, and (iii) Interactions Ratio.
For each of these high-level feature, we obtain a time series
\[ v_{1},v_{2},...,v_{\delta/30},\]
on which we compute a set of well-known statistical measures that represent the evolution over time of the series~\cite{wiens2012patient}. Namely, these statistical measures are the \textit{mean}, \textit{standard deviation}, \textit{linear weighted mean} and \textit{quadratic weighted mean},  \textit{average absolute change}, and \textit{maximum} of the series. These measures capture the evolution of the time series up to the current time and, especially, do not require us to know the whole time series, an important property for our Early Stage classification experiment.
We decided to derive time-series only for the three high-level features listed above. Indeed, we argue that the other features either have no temporal properties (e.g., lifetime, time to reach 90\% interactions), evolve in similar or predictable ways (e.g., diameter), or describe behaviors that are already captured by the selected time series. Moreover, evolution properties of time series would be especially used in Early Stage classification. Social networks need to perform feature extraction on this scenario at every time step: features derived from topological properties are too computationally expensive (e.g., diameter calculation runs in more than quadratic complexity w.r.t. the number of vertices) to be continuously calculated for each new post.
\section{Results}
\label{sec:results}
In this section we present the results of our experiments. In particular, we first discuss evaluation metrics and baseline (Section~\ref{sec:metrics}). We then report the results on the Early Stage Classification scenario (Section~\ref{sec:early_stages}), and the results on the Final Stage Classification scenario (Section~\ref{sec:final_stage}). 

\subsection{Evaluation Metrics}\label{sec:metrics}
As usual in binary classification, the classification baseline is the performance of a random classifier on the data: without any information regarding the propagation graph, it guesses either \textit{science} or \textit{conspiracy} with equal probability, as a coin flip.
The goal of our experiments is to show that structural features do not help more sophisticated models in improving the baseline performance.

Unfortunately, our dataset is highly imbalanced (composed by 89491 news for \textit{conspiracy}, and only 22650 news for \textit{science}).
With such imbalance, standard evaluation metrics (such as \textit{precision}, \textit{accuracy}, and \textit{recall}) can be misleading, because they do not account for the uneven class frequencies.
Even 
computing averages of these metrics using weights based on the class frequencies 
does not fit our intentions of consistently comparing our results with the fixed baseline.

To deal with this imbalance, we performed two distinct experiments: 
(i) we consider only metrics that take imbalance into account and use the full dataset, and (ii) we consider meaningful metrics with balanced dataset, obtained  \textit{undersampling} the original dataset.

To perform (i), as metrics we use Area Under Receiver Operating Curve (AUC)~\cite{hanley1982meaning}, and Cohen's Kappa~\cite{cohen1968weighted} (scaled into the interval [0,~1]). Indeed, the value of these metrics for a random classifier is exactly $0.5$, which we use as baseline. In this way, we can use the full dataset and still be able to compare our results with the baseline.

To perform (ii), we undersampled the most frequent class (i.e., \textit{conspiracy}). We therefore extracted exactly 22650 \textit{conspiracy} samples from the full dataset, and created a subsample with perfectly balanced classes.
To account for possible biases caused by the undersampling, we repeated the process several times, and averaged the outcomes.
Using perfectly balanced datasets, we can evaluate \textit{precision}, \textit{recall}, \textit{accuracy}, and \textit{$F_1$-score} values of our classifiers. Indeed, the value of these metrics for a random classifier on balanced data is exactly $0.5$, which we use as baseline.

Hereafter, if not differently specified, the metrics that require a positive class (such as \textit{precision}, \textit{recall}, and \textit{$F_1$-score}) use \textit{conspiracy} as the positive class, and \textit{science} as the negative one.

\subsection{Early Stage Classification}\label{sec:early_stages}
To evaluate this scenario, we recall that we used a total of 18 features. We then followed the methodology explained in Section~\ref{sec:experimental_design}, and evaluated three separate classifiers (namely, Linear Discriminant (LD), Random Forest (RF), Multi-Layer Perceptron (MLP)) at every 30 minutes time step. As discussed, we first evaluated AUC and Cohen's Kappa of these classifiers on the full, unbalanced dataset. We then evaluated precision, recall, and accuracy of these classifiers on undersampled balanced datasets.

\begin{figure}[htpb]
    \centering
    \includegraphics[width=0.45\textwidth]{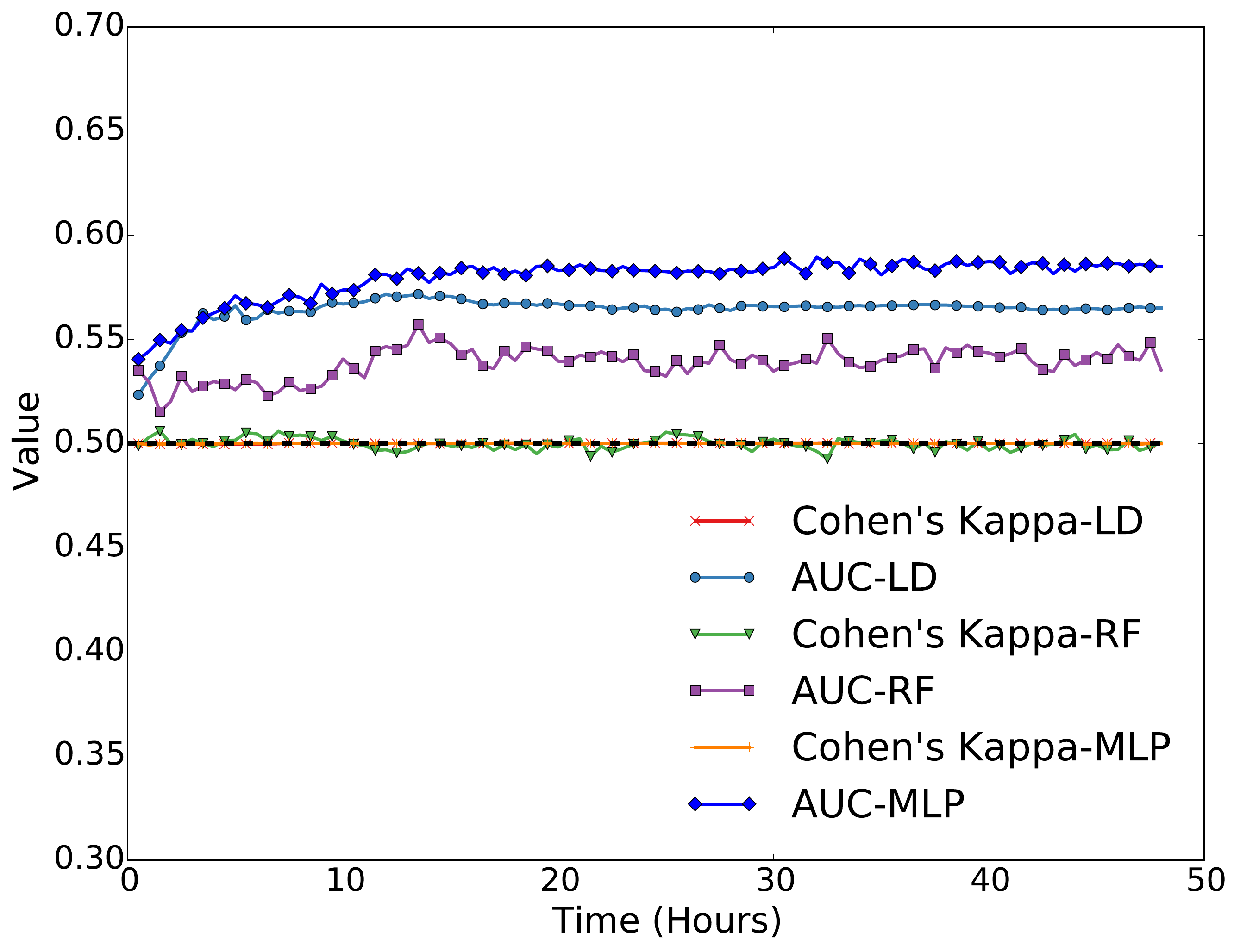}
    \caption{Early Stage Classification scenario, AUC and Cohen's Kappa, as a function of elapsed time.}\label{fig:early_stages_auc_cohen}
\end{figure}

Figure~\ref{fig:early_stages_auc_cohen} reports the AUC and the Cohen's Kappa metrics, on a 5-fold cross-validation scheme, at different time steps after the original post is published.
We can see that evolution properties of the propagation graph do not help significantly our classifiers. Indeed, the curves are close to the random classification baseline (dotted horizontal line), and far from perfect classification (i.e., 1.0).
Cohen's Kappa lies almost exactly on the random classification baseline, suggesting that the classification performance is almost random.
Using AUC, it is possible to set good classification thresholds, slightly improving the performance. MLP performs slightly better than RF and LD, and we can observe that results do not change significantly after a few hours of analysis. However the outcome remains unsatisfying (i.e., AUC under 0.6 for each classifier at each time step).

\begin{figure}[htpb]
    \centering
    \includegraphics[width=0.45\textwidth]{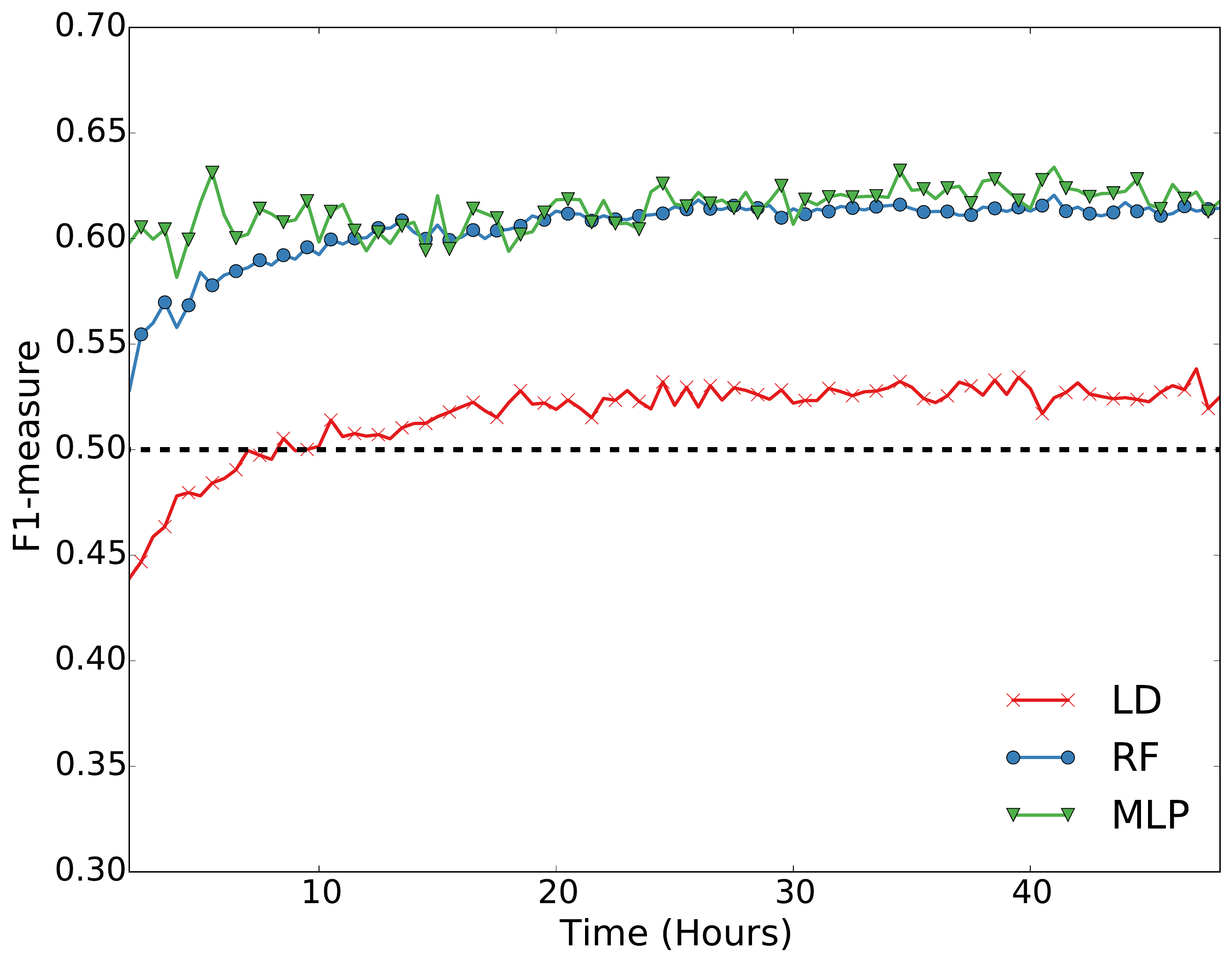}
    \caption{Early Stage Classification scenario, classification $F_1$-score, as a function of elapsed time.}\label{fig:early_stages_f1}
\end{figure}

Figure~\ref{fig:early_stages_f1} reports the \textit{$F_1$-score} obtained on the undersampled balanced dataset, on a 5-fold cross-validation scheme, averaged over ten repetitions of the undersampling procedure afterwards.
Figure~\ref{fig:early_stages_f1} further confirms the findings highlighted in Figure~\ref{fig:early_stages_auc_cohen}: classifiers are not able to leverage the evolution properties to discriminate between \textit{science} and \textit{conspiracy}.
In fact the \textit{$F_1$-score} curve, after a slight increment during the first 10 hours, stabilizes relatively close to the baseline, and remains below 0.65 throughout the whole 48 hours timespan.

\subsection{Final Stage Classification}\label{sec:final_stage}
To evaluate this scenario, we recall that we used a total of 28 features with three distinct classifiers: Linear Discriminant (LD), Random Forest (RF), Multi-Layer Perceptron (MLP). Again, we first evaluated the unbalanced dataset, and then evaluated the undersampled balanced datasets.

\begin{figure}[htpb]
    \centering
    \includegraphics[width=0.45\textwidth]{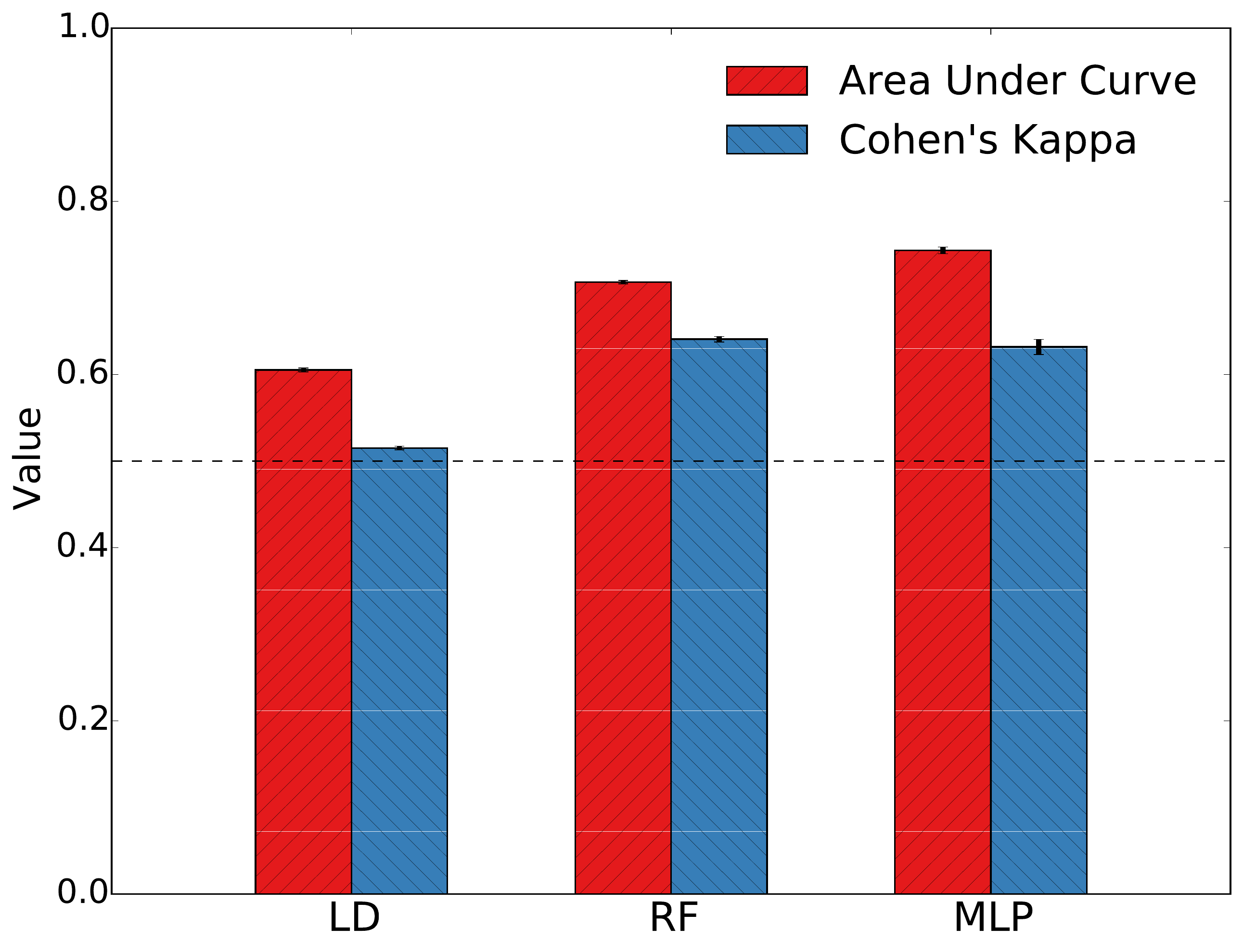}
    \caption{Final Stage Classification scenario, AUC and Cohen's Kappa.}\label{fig:auc_cohen}
\end{figure}

Figure~\ref{fig:auc_cohen} reports average AUC and Cohen's Kappa on a 5-fold cross-validation scheme, on the full dataset. The dotted horizontal line shows the baseline performance of random classification. Indeed, we observe that our classifiers do not significatively improve the baseline, with the metrics remaining below $0.75$.
LD performance is poorer than RF and MLP, probably due to the simplicity of the classification tool.

\begin{figure}[htpb]
    \centering
    \includegraphics[width=0.45\textwidth]{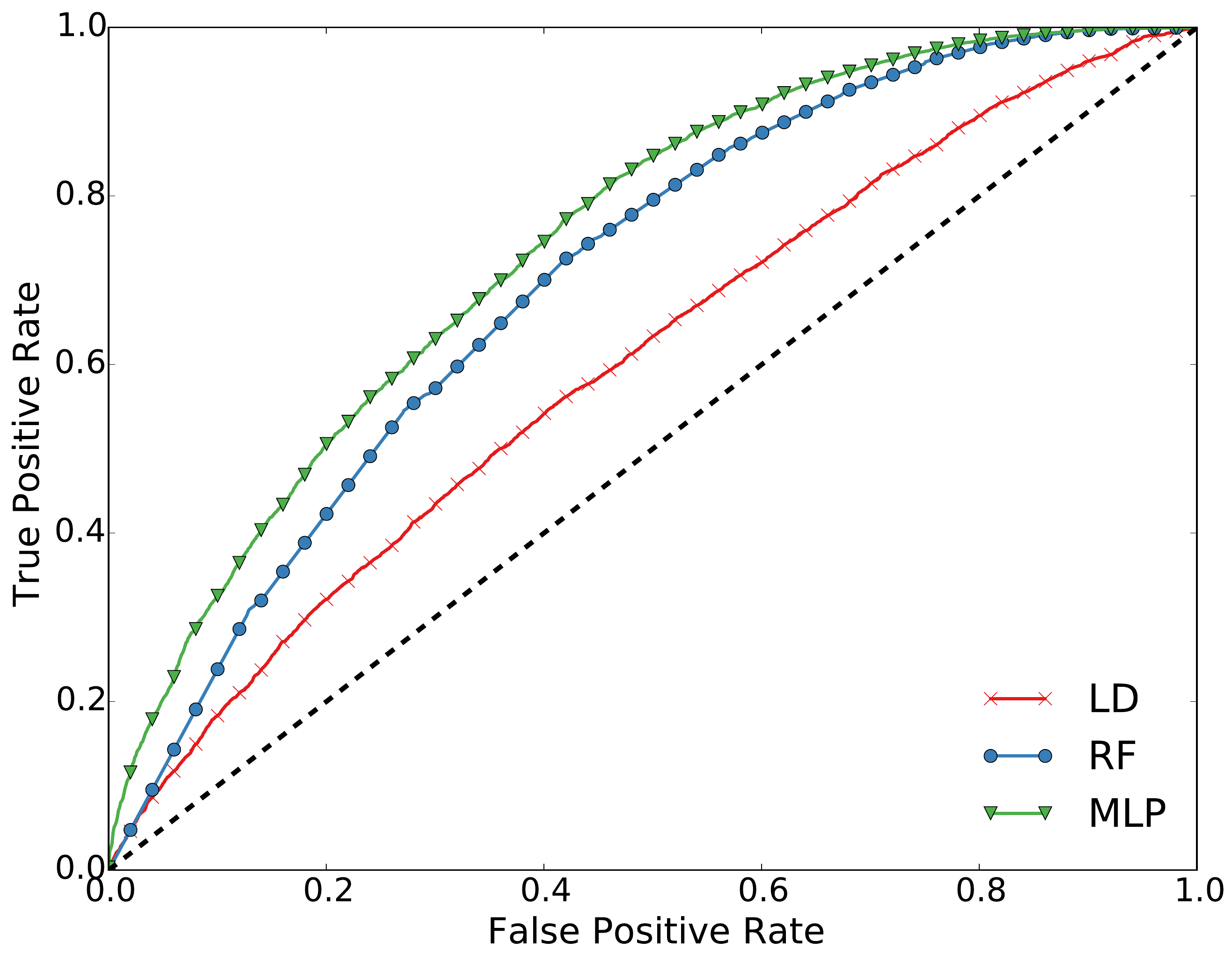}
    \caption{Final Stage Classification scenario, ROC curves. For visual clarity, we linearly interpolated the points in the curves, and plotted markers at each 0.02 step in the false positive rate.}\label{fig:roc_analysis}
\end{figure}

In Figure~\ref{fig:roc_analysis} we report the ROC curve, on a 5-fold cross-validation scheme, on the full dataset.
From Figure~\ref{fig:roc_analysis} we can observe that no classifier can reach a good tradeoff between true positive rate and false positive rate.
Indeed, the curves are relatively close to the baseline (diagonal dotted line), meaning that, as the decision threshold changes, lots of samples are misclassified.

Table~\ref{tab:final_stage_us} reports several metrics computed on the undersampled balanced dataset.
Results are averaged on a 5-fold cross-validation scheme, and then over ten repetitions of the undersampling procedure.
These metrics show that even if the classifiers are able to improve the baseline slightly, they can not reach good performance.
{
\tabulinesep = 1.5mm
\begin{table}[htpb]
\centering
\begin{tabu} to \linewidth {X@{\qquad}X@{$\;\;$}X@{$\;$}X@{$\;\;$}X}
\hline 
\textbf{Classifier} & \textbf{Precision} & \textbf{Recall} & \textbf{Accuracy} & \textbf{$F_1$ score}  \\
\hline\hline 
 LD & 0.578 & 0.523 & 0.570 & 0.549\\
 RF & 0.654 & 0.742 & 0.675 & 0.695\\
 MLP & 0.659 & 0.688 & 0.665 & 0.672\\
\hline

\end{tabu}
\caption{Final Stage Classification scenario, performance of the classifiers.}
\label{tab:final_stage_us}
\end{table}
}


\section{Conclusions}
\label{sec:conclusions}
Early detection of misinformation plays a crucial role in social networks. 
In this paper, we analyzed the difficulty of discerning conspiracy posts from scientific posts on Facebook. We focused on using only structural features of content propagation, because they cannot be easily manipulated by misinformation creators.
Our results show that misinformation classification during its early propagation stage with these features is unsuccessful, suggesting that the spreading dynamics captured by our features are independent on the type of content.
Furthermore, even considering the cascade at the end of content propagation does not help: also in this case, the improvement provided by a classifier over random coin flips is negligible.

Our findings suggest that in Facebook users interact with different types of content in similar ways, reinforcing the hypothesis of echo chambers~\cite{del2016spreading}. Inside these chambers, strongly polarized by topic~\cite{bessi2015viral}, content propagation exhibits very similar structural properties, that are therefore less useful in content classification.
These results highlight the necessity of including content-related features, or polarization metrics, in future analysis (i.e., whether particular users and their echo chambers are more polarized towards one type of content).
Unfortunately, misinformation creators can easily control content-related features, in order to avoid algorithmic detection. Moreover, user polarization can be clearly understood from past users' behaviors, but it takes time to understand polarization of new users.
Hence, automatic detection of fake news remains an open challenge.

\paragraph{Future Work.}
The employed dataset  has some limitations, bound to the Facebook API: (i) it only contains Facebook public information; (ii) it does not contain the timestamp of likes, one of the most common interactions; and (iii) it does not always provide information about whether interaction with content happened because of interactions of user's friends. We would like to analyze finer-grained data, that takes these factors into account, because it could lead to improved results.

In our analysis, we identified a set of features and used them in well-known classifiers. Our experiments were extensive, but not complete. However, we expect that the use of different models would not provide significant improvement. This claim needs to be further validated, possibly also using more recent datasets.

In the future, we also plan to test different methods for misinformation classification, based on user polarization and content-related features, to investigate whether these information could help propagation properties, and overcome the difficulty of this problem.

\bibliography{bibliography}
\bibliographystyle{aaai}
\end{document}